\documentclass[journal,10pt]{IEEEtran}

\usepackage{url}
\usepackage{subfigure}
\usepackage{graphicx}

\usepackage{hyperref}
\hypersetup{colorlinks, citecolor=green, filecolor=pink, linkcolor=red, urlcolor=blue }

\def\ie{\textit{i.e.}}

\def\eg{\textit{e.g.}}

\begin{document}


\title{Blockchain Intelligence: When Blockchain Meets Artificial Intelligence}

\author{Zibin Zheng,~\IEEEmembership{Senior Member,~IEEE}, Hong-Ning Dai,~\IEEEmembership{Senior Member,~IEEE}, Jiajing Wu,~\IEEEmembership{ Member,~IEEE} 
\thanks{Z. Zheng, J. Wu are with School of Data and Computer Science, Sun Yat-sen University, China (email: zhzibin@mail.sysu.edu.cn, wujiajing@mail.sysu.edu.cn).}
\thanks{H.-N. Dai is with Faculty of Information Technology, Macau University of Science and Technology, Macau (email: hndai@ieee.org).}
}

\maketitle

\begin{abstract}
Blockchain is gaining extensive attention due to its provision of secure and decentralized resource sharing manner. However, the incumbent blockchain systems also suffer from a number of challenges in operational maintenance, quality assurance of smart contracts and malicious behaviour detection of blockchain data. The recent advances in artificial intelligence bring the opportunities in overcoming the above challenges. The integration of blockchain with artificial intelligence can be beneficial to enhance current blockchain systems. This article presents an introduction of the convergence of blockchain and artificial intelligence (namely blockchain intelligence). This paper also gives a case study to further demonstrate the feasibility of blockchain intelligence and point out the future directions.
\end{abstract}

\begin{keywords}
Blockchain; Artificial Intelligence; Smart Contract; Machine Learning
\end{keywords}

\section{Introduction}

Blockchain has received extensive attention recently due to its provision of secure data sharing services with traceability, immutability and non-repudiation. Despite the merits of blockchain, the development of blockchain technologies has undergone a number of challenges including poor scalability, difficulties in operational maintenance, detecting vulnerable codes in smart contracts and identifying malicious behaviours in blockchain historical data.

The recent advances in artificial intelligence (AI) have greatly propelled the evolution of diverse business applications. The integration of AI with blockchain has the potentials to overcome the limitations of blockchain. We name the intelligent capability bestowed by AI on the blockchain ecosystem as \emph{blockchain intelligence}. In particular, AI approaches such as machine learning, data mining and data visualization may help to capture the abnormal behaviours in blockchain, identify risks in transactions, detect possible vulnerable program codes in smart contracts, and so on. Consequently, proactive and autonomic actions can be made to prevent blockchain from disruptive or illegal actions, and thus make the blockchain more intelligent.

\begin{figure*}[t]
\centering
		\includegraphics[width=16cm]{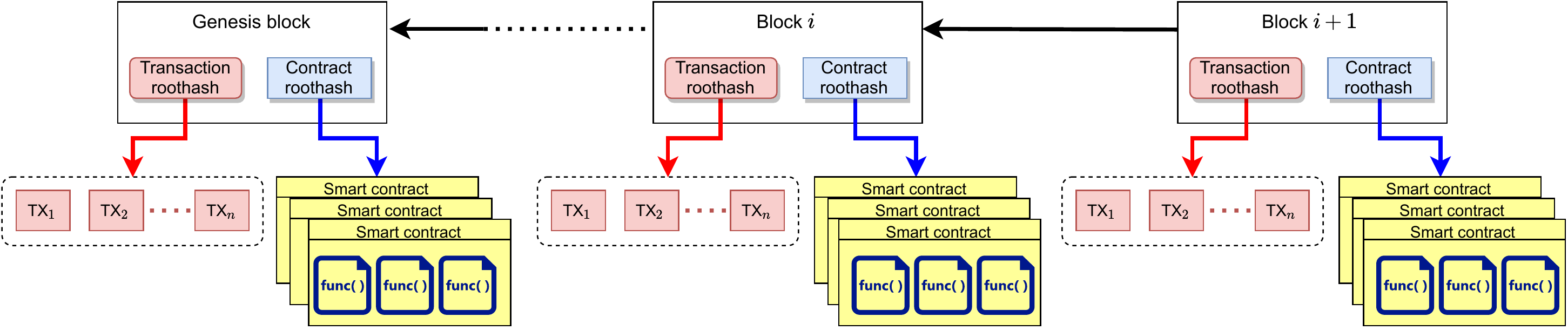}
		\caption{Overview of blockchain}
		\label{fig:blockchain}
\end{figure*}

This article aims at reviewing blockchain technologies as well as AI technologies, presenting an in-depth analysis on integrating blockchain with AI, providing implications of enabling technologies of blockchain intelligence. In contrast to most of recent studies on the integration of blockchain and AI (which mainly focuses on applying blockchain to overcome shortcomings of AI, such as security and privacy vulnerabilities~\cite{Dinh:Computer18}), this article mainly concentrates on improving blockchain systems by applying AI technologies. In summary, the main contributions of this article are summarized as follows:
\begin{itemize}
    \item We first give an overview of blockchain technology and point out the challenges in existing blockchain systems.
    \item We then review the advances in AI and formally introduce the convergence of AI and blockchain followed by a discussion on opportunities brought by blockchain intelligence.
    \item We next present a case study to demonstrate the feasibility of blockchain intelligence.
\end{itemize}

\section{Overview of blockchain technologies}
\label{sec:blockchain}

As a disruptive software technology, blockchain is reshaping diverse business sectors. Blockchain is essentially a chain-like data structure storing transactions verified by majority of nodes throughout the whole network as shown in Fig.~\ref{fig:blockchain}. Since the committed transactions in the blockchain have been stored at every node, they are extremely difficult to be altered or falsified. Integrating with digital signature and asymmetric encryption, blockchain data is authenticated and auditable, implying non-repudiation of the transaction initiator. The length of blockchain keeps growing with the new validated transaction being appended at end of the chain. Data analytics on blockchain data can potentially extract valuable information. 

The development of blockchain technologies has experienced two phases: 1) blockchain 1.0 (\ie, symbolized by digital currency) and 2) blockchain 2.0 (\ie, symbolized by smart contracts)~\cite{li2017survey}. In blockchain 1.0, blockchain has been mainly used for digital currencies like Bitcoin. The appearance of blockchain has promoted the development of smart contracts. A smart contract essentially consists of a number of computerized contractual agreements consented by multiple parties~\cite{szabo1997idea}. The contractual clauses embedded in smart contracts will be triggered and automatically executed when a certain condition is satisfied (\eg, who breaches the contract will be automatically imposed with a fine).

Smart contracts have been implemented on top of blockchain as shown in Fig.~\ref{fig:blockchain}. The approved contractual clauses are converted into executable computer programs. The logical connections between  contractual clauses have also been preserved in the form of logical flows in programs (\eg, \texttt{if-else-if} statement). The execution of each contract statement is recorded as an immutable transaction stored in the blockchain. Meanwhile, smart contracts guarantee appropriate access control and contract enforcement. In particular, developers can assign access permissions for each function in the contract. 

Although blockchain and blockchain-enabled smart contracts are promising in reshaping various industrial sectors, the intrinsic limitations of blockchain systems also lead to the following challenges.

\emph{1) Operational maintenance}. Due to the decentralization and heterogeneity of blockchain systems, it is difficult to identify the potential factors affecting the performance of blockchain. For example, the \emph{transaction throughput} bottleneck of Hyperledger Fabric is different from that of Bitcoin and Ethereum since different consensus algorithms are adopted. Moreover, like other software systems, smart contracts consist of a number of computer programs, which may suffer from software bugs, malicious codes and incompatibility of running environments. Consequently, it is crucial to achieve the intelligent and robust operational maintenance of complex blockchain systems.
    
\begin{figure*}[t]
	\centering
		\includegraphics[width=16cm]{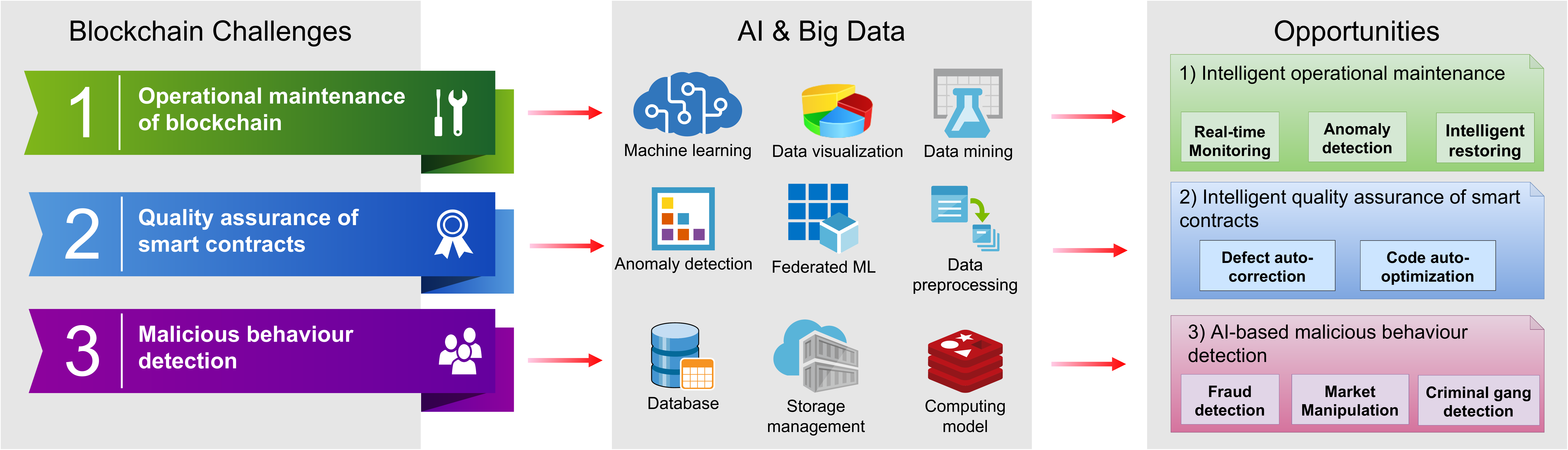}
		\caption{Opportunities brought by artificial intelligence to address challenges of blockchain}
		\label{fig:ai}
\end{figure*}
	
\emph{2) Quality assurance of smart contracts}. Smart contracts suffer from a number of software vulnerabilities such as {\it re-entrancy} vulnerability~\cite{li2017survey}, overcharging issue~\cite{chen2017under}, randomness controlling~\cite{bonneau2015bitcoin} and Decentralized Autonomous Organization (DAO) attack~\cite{thedaoattack}. In addition, contract correctness is also crucial to smart contracts since it is nearly impossible to make any revisions once they are deployed on top of blockchains. However, like software systems, smart contracts often contain programming bugs which may lead to crashes or misbehaviours while it is challenging to detect and identify these bugs due to the complexity of smart contracts.  
    
\emph{3) Malicious behaviour detection}. Besides legal businesses, blockchain may be exploited for malicious activities which are nevertheless difficult to be detected due to the \emph{pseudonymity} of blockchain (\ie, anonymous blockchain addresses). On the other hand, the encrypted blockchain data also leads to the difficulty of detecting and identifying malicious behaviours via simply data analytics. Moreover, the massive volume and heterogeneity of blockchain data as well as diversity of user behaviours make the problem even worse. As a result, conventional classification-based methods (\eg, machine learning methods) cannot be directly applied.

\section{Opportunities brought by Artificial Intelligence}

Artificial intelligence (AI) as a broad discipline covering machine learning and cognitive computing is an ability of intelligent agents conducting intellectual tasks. The recent advances in big data, AI technologies (such as deep neural networks) and general purpose computer hardware such as graphic processing units (GPUs) have greatly driven the development of AI. Consequently, we have witnessed the proliferation of diverse AI applications such as computer vision, natural language processing, speech recognition, sentimental analysis. Big data plays a critical role in propelling AI as well as AI applications. For example, deep learning mainly based on deep neural networks (DNNs) has achieved superior performance thanks to the availability of massive data so that DNNs can extract (or learn) enough features from large volume of data. 

The appearance of diverse blockchain systems has generated the enormous volumes of blockchain data, which are publicly available to everyone. Take Bitcoin as an example. As reported by Statista (https://www.statista.com/) at the end of the  third quarter of 2019, Bitcoin contains nearly 242 GB data. Data analytics on the massive blockchain data cannot only extract huge business values but also bring great opportunities to overcome the aforementioned challenges of blockchain systems. The recent advances in AI have also greatly driven the development of big data analytics~\cite{dai:2019}. Thus, the integration of AI and blockchain technologies can potentially overcome the aforementioned challenges of blockchain systems, thereby forming intelligent blockchain systems. We name such integration of blockchain with AI as \emph{blockchain intelligence}. It is worth mentioning that there are several research efforts~\cite{{Dinh:Computer18,Salah:WIDM20}} on integrating blockchain with AI while most of them are mainly concentrated on exploiting blockchain for AI so as to overcome the emerging security and privacy vulnerabilities of AI. In contrast to these studies, our work in this paper is mainly focused on solving the intrinsic issues of blockchains by using AI technologies.

We summarize the opportunities brought by AI to enhance incumbent blockchain systems as follows (as shown in Fig.~\ref{fig:ai}).

\emph{1) Intelligent operational maintenance of blockchain.} Blockchain generates huge amount of data in a real-time manner. Analyzing blockchain data, we can detect the possible faults, forecast the failures and identify the performance bottleneck so as to tune or adjust the performance of blockchain systems. There are four different levels of data analytics including: descriptive analytics, diagnostic analytics, predictive analytics and prescriptive analytics. In particular, the work of~\cite{Dinh:TKDE18} presents a common platform to evaluate the performance of three representative blockchain systems: Ethereum, Parity and Hyperledger Fabric via descriptively analyzing blockchain data. Meanwhile, the descriptive analytics of blockchain log data can help to monitor the real-time performance of blockchain systems and identify the possible faults~\cite{PZheng:ICSE18}. In addition to diagnostic analytics on blockchain data, predictive analytics is also necessary to anticipate the performance bottleneck of blockchain systems. Unlike diagnostic and predictive analytics, prescriptive analytics can simulate and optimize blockchain systems so as to improve the reliability of blockchain systems.

\emph{2) Intelligent quality assurance of smart contracts.} Like computer software, smart contracts may contain bugs or faulty programming codes, which are vulnerable to crashes and malicious attacks. It is crucial to detect and identify bugs in smart contracts so as to achieve the ultimate goal of intelligent quality assurance of smart contracts. The work of~\cite{Luu:CCS16} presents a symbolic-execution platform namely Oyente to detect and recognize potential bugs. Meanwhile, smart contracts are essentially program codes that are sensitive to execution cost (e.g., the execution of smart contracts is charged by gas in Ethereum). Thus, it is a necessity to identify the gas-costly patterns and correct these vulnerable smart contracts. In~\cite{chen2017under}, Chen et al. developed a tool namely GASPER to identify and locate seven gas-costly patterns via analyzing bytecodes of Ethereum smart contracts. Machine learning methods can be used to detect and recognize vulnerable bugs in smart contracts automatically. Moreover, the growing number of smart contracts also brings the opportunities to automate the composition of multiple contracts. In particular, we can find and identify contracts that fulfill users' requirements and composite them together to realize more comprehensive applications. Different from conventional distributed software systems such as web services~\cite{Rodriguez-Mier:TSC16}, smart contracts lack of semantic description and Quality-of-Service (QoS) evaluation metrics. As a result, new AI-based approaches are expected to automate labeling semantics of smart contracts and offer data-driven QoS evaluation of smart contracts.

\emph{3) Automated malicious behaviour detection.} The decentralized blockchain systems result in the difficulty in auditing malicious behaviours such as money laundering, phishing, gambling and scams that occurred in blockchain platforms. Blockchain systems have generated massive transaction data, which are essentially available to everyone, whereas the historical transaction data are pseudonymous through anonymizing account addresses. The massive blockchain data brings the opportunities in auditing and detecting malicious behaviours. Big data analytics on massive blockchain data can help to identify malicious users, recognize behaviour pattern, analyze market manipulation, detect scams. The work of~\cite{TChen:INFOCOM2018} presents a cross-graph analysis on Ethereum data and identify several major activities occurring on Ethereum blockchain platforms. As in~\cite{WChen:WWW18}, a machine learning based approach was proposed to detect and capture Ponzi schemes that took place in Ethereum. The work of~\cite{WChen:INFOCOM2019} analyzes the leaked transaction history of Mt. Gox Bitcoin exchange and identify a number of market manipulation patterns via singular value decomposition (SVD) method. Moreover, malicious users may exploit multiple anonymous accounts to form a criminal gang to conduct illegal activities on blockchain systems. New machine learning approaches as well as association analysis on multiple accounts are expected to address this issue.
    
The advances in AI, machine learning and big data analytics bring numerous opportunities to address the aforementioned blockchain challenges. We next present a case study to demonstrate the feasibility of blockchain intelligence.

\section{Case study}


Big data analytics of blockchain data is beneficial to fraud recognition of transactions and vulnerability detection of smart contracts. However, it is also challenging to conduct big data analytics of blockchain data. 1) It is extremely time consuming to download the entire blockchain data due to the bulky blockchain size, e.g., it took more than one week and over 500 GB storage space to fully synchronize (\ie, download) the entire Ethereum at a newly-joined peer. 2) It requires substantial efforts in extracting and processing blockchain data. First, blockchain data is stored at clients in heterogeneous and complex data structures, which cannot be directly analyzed. Meanwhile, the underlying blockchain data is either binary or encrypted. Thus, it is a necessity to extract and process binary and encrypted blockchain data so as to obtain valuable information while this process is non-trivial as conventional data analytic methods may not work for this type of data. 3) There is no general data extract tool for blockchain data. Although several open source tools for blockchain data extraction are available, most of them can only support to extract partial blockchain data (not the entire data).

\subsection{Data Extraction from Ethereum}

\begin{figure}[t]
    \centering
    \includegraphics[width=8.2cm]{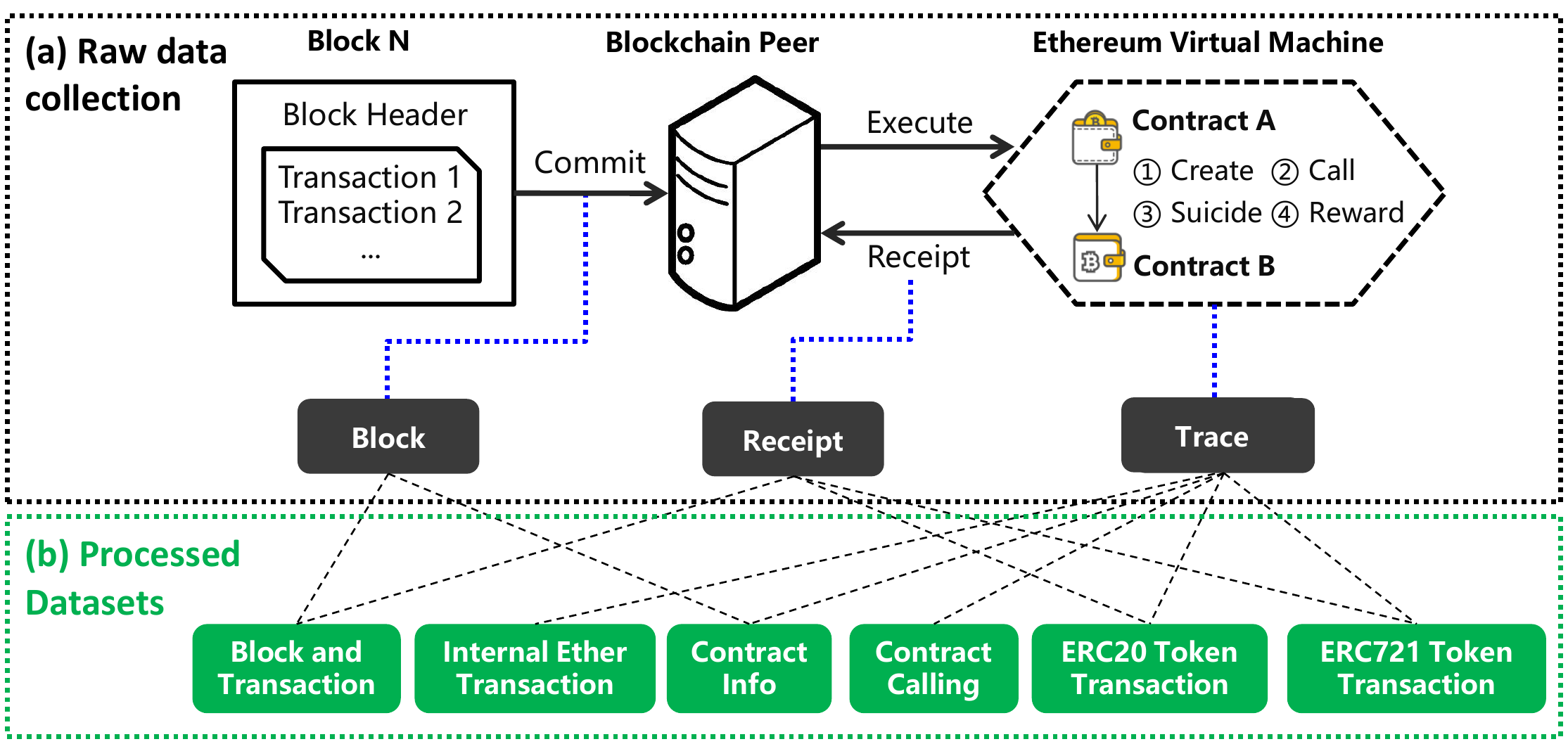}
    \caption{Data processing during Ethereum transaction flow}
    \label{fig:dataprocess}
\end{figure}

To address the above challenges, we propose a blockchain data analytics framework namely XBlock-ETH to analyze Ethereum data. In particular, we extract raw data consisting of 8,100,000 blocks of Ethereum. Fig.~\ref{fig:dataprocess}(a) illustrates the typical Ethereum transaction execution flow from Block $N$ to EVM through blockchain peer. During this procedure, we collect the three types of blockchain raw data: Block, Receipt and Trace. Since the analysis on the raw blockchain data is difficult, we process and categorize the obtained Ethereum Blockchain data into six datasets: \textit{(1) Block and Transaction}, \textit{(2) Internal Ether Transaction}, \textit{(3) Contract Information}, \textit{(4) Contract Calls}, \textit{(5) ERC20 Token Transactions}, \textit{(6) ERC721 Token Transactions} as shown in Fig.~\ref{fig:dataprocess}(b). It is non-trivial to process the raw since it requires substantial efforts in extracting useful information from raw data and associating with six datasets. 

\begin{figure}[t]
\subfigure[Macro view of GasPrice]{
\centering
\includegraphics[width=4.1cm]{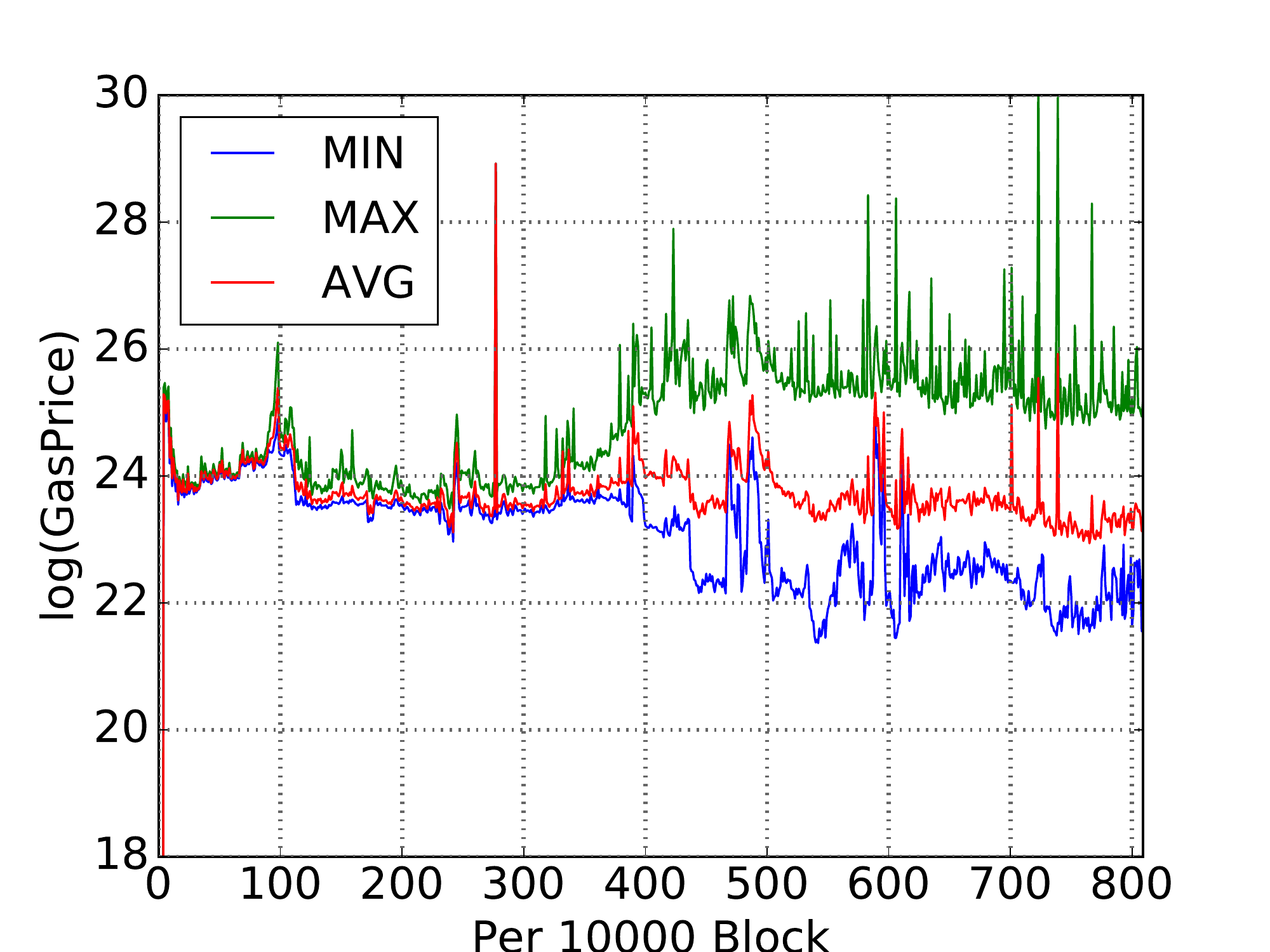}\label{1_gasprice}
} 
\subfigure[Micro view of GasPrice]{
\centering
\includegraphics[width=4.1cm]{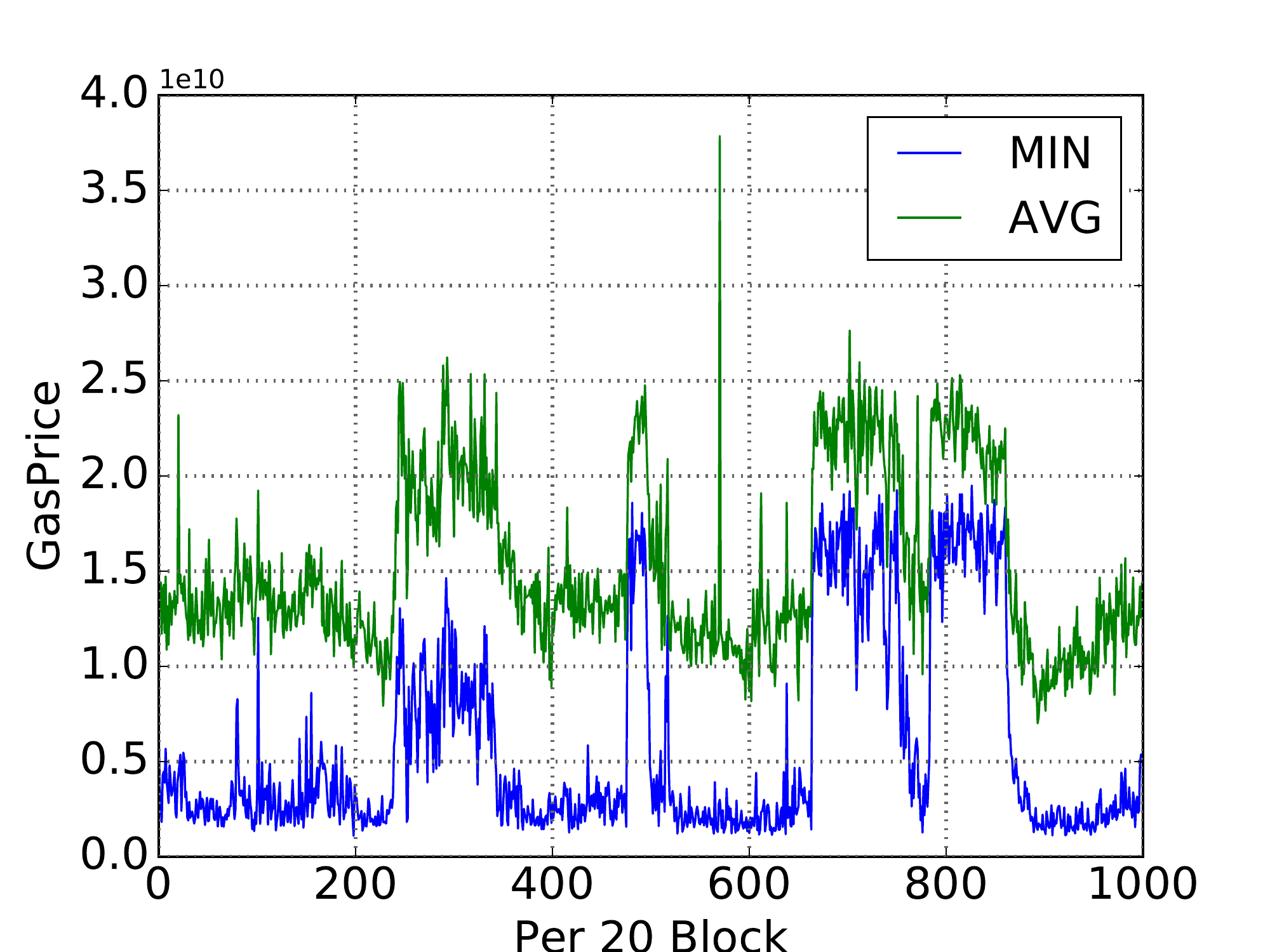}\label{1_gaspriceOneday}
}
\caption{Data visualization of Gas Price of Ethereum} 
\label{fig:cost}
\end{figure}

We then conduct statistic analysis on these refined datasets. In Ethereum, a miner has a higher priority to package the transactions with higher ``\texttt{gasPrice}'' into the block. The visualization of ``\texttt{gasPrice}'' is shown in Fig.~\ref{fig:cost}. In a macro view (as shown in Fig.~\ref{1_gasprice}), the ``\texttt{gasPrice}'' is gradually decreasing with the development of the Ethereum community, except for several peaks caused by extremely frequent transaction when the network is congested. In a micro view (as shown in Fig.~\ref{1_gaspriceOneday}), we extract the time from 8,000,000 to 8,020,000 blocks and find that such fluctuations of ``\texttt{gasPrice}'' can be observed by the tidal law. This observation implies that the fluctuations of ``\texttt{gasPrice}'' can potentially be predicted.

\subsection{Detection Ponzi Schemes from Ethereum}

\begin{figure*}[t]

\subfigure[LooneyLottery contract (\ie, normal contract)]{
\centering
\includegraphics[width=8.1cm]{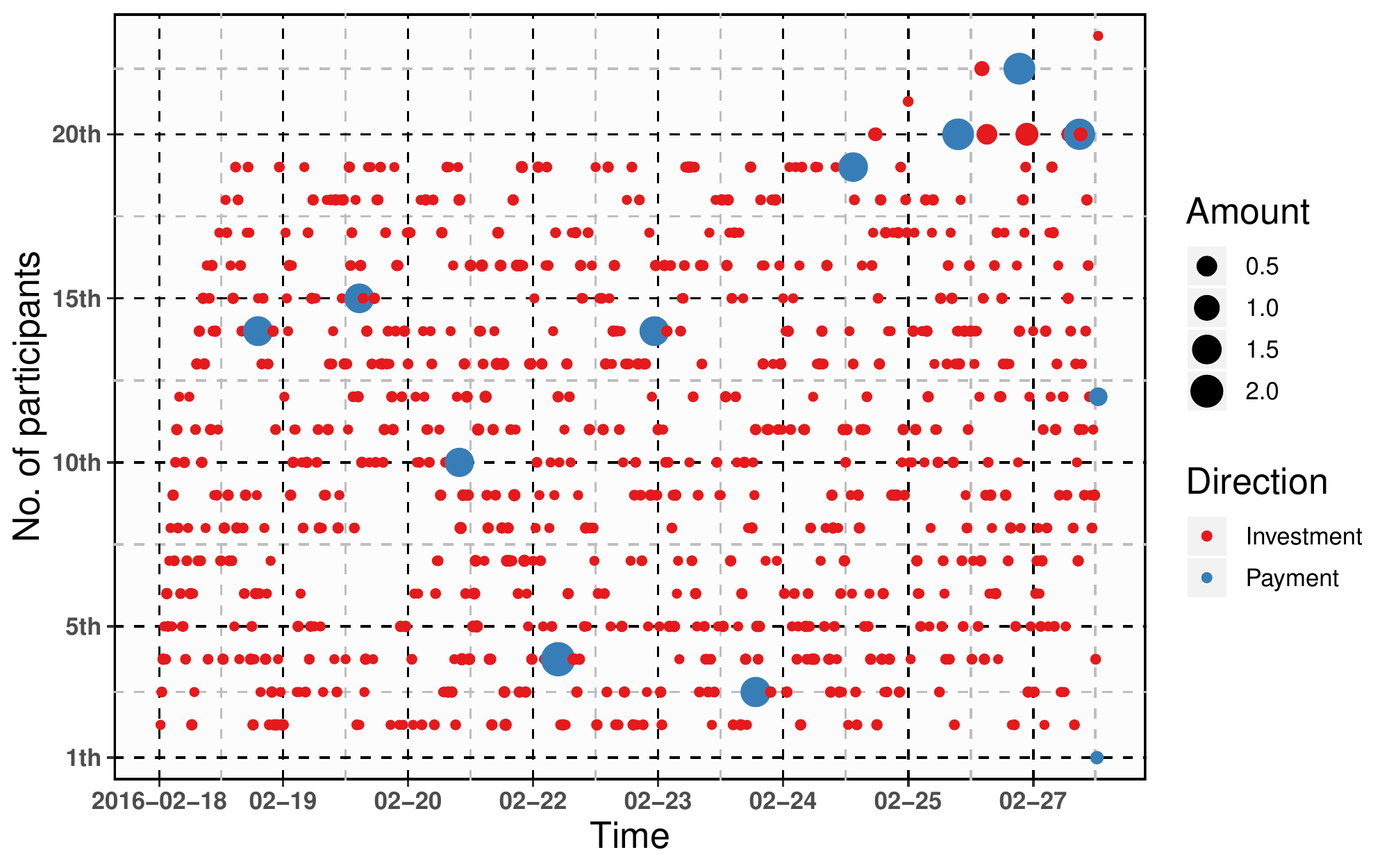}\label{lottery}
} 
\subfigure[Rubixi contract (\ie, Ponzi scam contract)]{
\centering
\includegraphics[width=8.1cm]{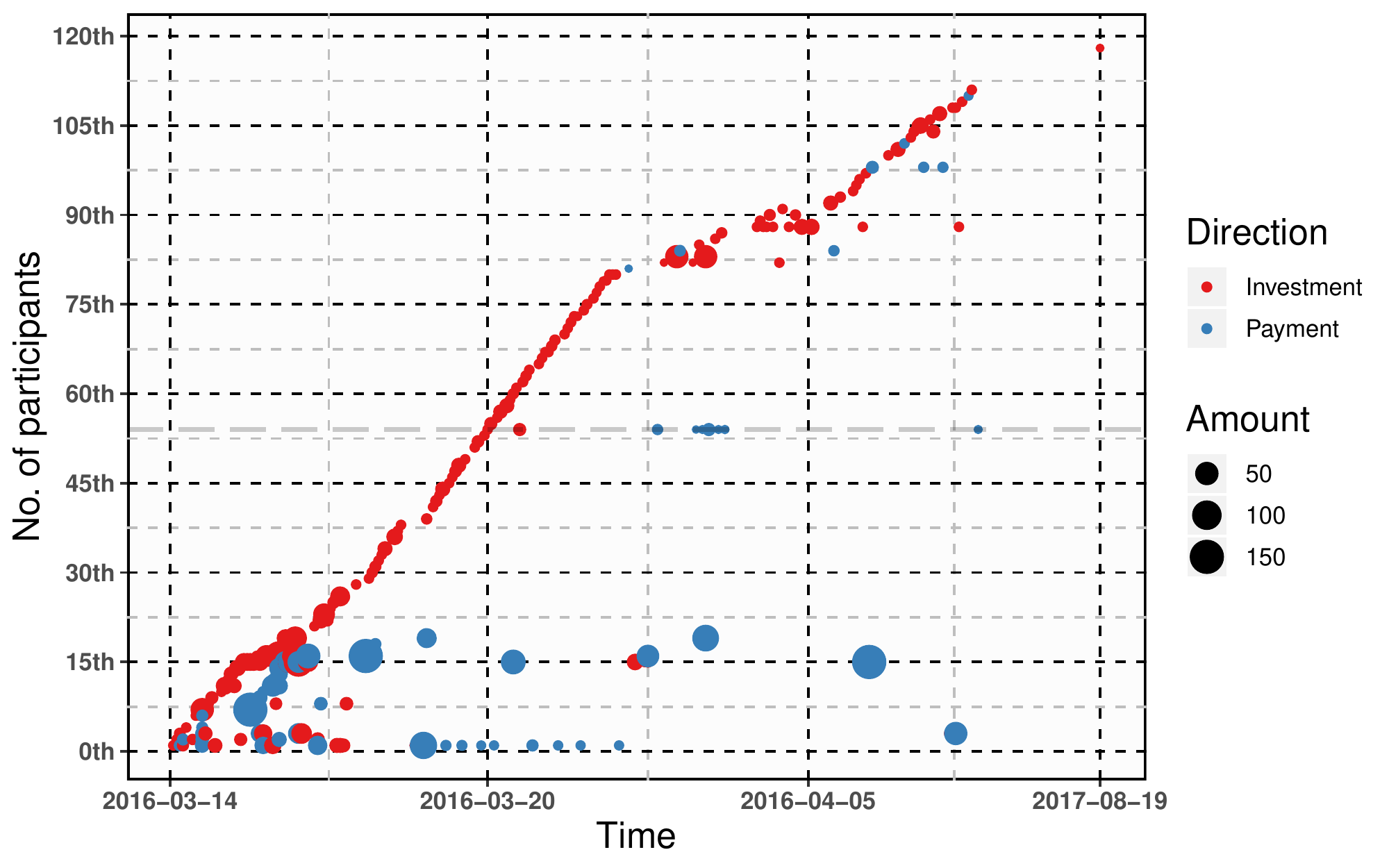}\label{rubixi}
}
\caption{Ether Flow Graph of two smart contracts~\cite{WChen:WWW18}.} 
\label{fig:ponzi}
\end{figure*}

Blockchain can also be exploited to conduct illegal activities such as scams. For example, in~\cite{WChen:WWW18}, we propose a method to detect Ponzi scams in Ethereum blockchain through extracting and analyzing key characteristics of user accounts and operation codes in Ethereum contracts. Fig.~\ref{fig:ponzi} makes a comparison between a normal contract and a Ponzi scheme contract in terms of Ether Flow Graph, where the horizontal axis denotes the time line, the vertical axis represents the number of participants in a particular contract, red circles and blue circles denote the investment transactions and payment transactions, respectively. In addition, the size of circle also represents the amount of transactions, \ie, the larger circle means the larger amount of transactions.

Fig.~\ref{lottery} is a normal lottery contract while Fig.~\ref{rubixi} shows a typical Ponzi scam (namely Rubixi). We can observe several significant differences between Fig.~\ref{lottery} and Fig.~\ref{rubixi}: 1) there are more participants in Fig.~\ref{rubixi} than Fig.~\ref{lottery}; 2) there are more payment transactions in Fig.~\ref{rubixi} than Fig.~\ref{lottery} which exhibits more randomness in the number of transactions. After extracting the key features and applying other data mining and machine learning methods, we can successfully classify Ponzi scams from other normal activities.

\section{Conclusion and future directions}

This article first reviews the blockchain technologies and analyze the challenges in blockchain systems. We then introduce artificial intelligence (AI) as well as opportunities brought by AI to blockchain systems. We name such integration of blockchain and AI as \emph{blockchain intelligence}. We mainly discuss that AI bring benefits to blockchain in aspects of \emph{intelligent operational maintenance of blockchain}, \emph{intelligent quality assurance of smart contracts} and \emph{automated malicious behaviour detection}. In addition, we also give a case study to further demonstrate great potentials of blockchain intelligence. 

We believe that the integration of AI with blockchain technology will further drive the benignant development of blockchain systems. We outline the future directions in blockchain intelligence as follows.
\begin{itemize}
    \item \emph{Real-time and automated operational maintenance of blockchain.} Instead of performance monitoring and fault detection, the future intelligent operational blockchain systems are expected to conduct real-time monitoring on multiple performance metrics of blockchain systems and achieve the automated restoring from crash spots.
    
    \item \emph{Collective intelligence bestowing smart contracts.} The current QoS assurance approaches of smart contracts are mainly based on the analysis on contract patterns (\ie, detecting and classifying vulnerable contracts) at a sole peer in the blockchain. In contrast to a single intelligent agent, \emph{collective intelligence} can motivate all participants to engage in analyzing and reasoning, thereby contributing their collective knowledge to make better decision in a global context. In the future, collective intelligence is expected to integrate with decentralized blockchain system so as to offer a trustworthy and intelligent provision of blockchain services.

    \item \emph{Integrating multiple machine learning approaches to monitor and supervise blockchain data.} Decentralization of blockchain brings the difficulty in monitoring and supervising transactions in blockchain platforms, consequently resulting in a number of illegal behaviours. Meanwhile, both heterogeneity and pseudonymity of blockchain data make this situation even worse. In the future, multiple machine learning approaches should be integrated together to extract key features from different types of blockchain data. In addition, a dynamic graph (or network) of transactions is also expected to identify the association between different accounts so as to recognize the malicious behaviours. 

\end{itemize}

\end{document}